\documentclass[
]{ceurart}

\usepackage{listings}
\usepackage[T1]{fontenc}
\usepackage{graphicx}
\usepackage{xstring}
\usepackage{longtable}
\usepackage{booktabs}
\usepackage{subcaption}

\begin{document}

\copyrightyear{2026}
\copyrightclause{Copyright for this paper by its authors.
  Use permitted under Creative Commons License Attribution 4.0
  International (CC BY 4.0).}

\conference{PEAF 2026: The First International Workshop on Pedagogical Evaluation of Automated Feedback, June 28 2026, Seoul, South Korea}

\title{Supporting Tutors in the Gig Economy with Automated Feedback: A Case Study on Ringle}


\author[1]{Yeon Su Park}[]
\fnmark[1]
\address[1]{KAIST, Daejeon, Republic of Korea}

\author[2]{Sieun Kim}[]
\fnmark[1,2]
\address[2]{University of Michigan, Ann Arbor, Michigan, USA}

\author[3]{Keighley Overbay}[]
\fnmark[3]
\address[3]{Samsung Research, Seoul, Republic of Korea}

\author[1]{Seoyoung Kim}[]

\author[4]{Sewook Wee}[]
\address[4]{Ringle, San Mateo, California, USA}

\author[1]{Daho Jung}[]

\author[1]{Juho Kim}[]

\fntext[1]{These authors contributed equally.}
\fntext[2]{Sieun conducted this work while she was at Seoul National University.}
\fntext[3]{Keighley conducted this work while she was at Ringle.}


\begin{abstract}
The rise of online tutoring platforms in the gig economy has made education more scalable, flexible, and on-demand. These platforms rely on learner evaluations as the primary feedback for tutors and platforms. However, such feedback offers limited guidance for tutors' improvement and makes it difficult to monitor tutor quality at scale. To this end, we explored AI-powered automated feedback and how tutors perceive and respond to it. We deployed a research probe on Ringle, a popular online English tutoring platform, that analyzed tutors' lessons and provided automated feedback. We then surveyed 36 tutors about their experience. Our findings reveal that while tutors perceived automated feedback more negatively than learner feedback, they found it useful for self-monitoring and understanding platform expectations, though discrepancies between them often caused confusion. Based on these insights, we propose design considerations for feedback systems for online educational gig platforms.
\end{abstract}


\maketitle

\section{Introduction}

The gig economy has expanded opportunities for flexible work across various fields. In education, online language tutoring platforms such as Preply, italki, Cambly, Verbling, and Ringle have become increasingly prevalent. Driven by global demand for English learning, these platforms let individuals tutor using their English proficiency, providing scalable and accessible learning~\cite{xia2022understanding}.

A central challenge of such platforms is maintaining worker quality at scale~\cite{wood2019good}. To address this, platforms commonly rely on user feedback such as ratings and reviews to evaluate performance and guide improvement. However, prior work shows that such feedback tends to be overly positive, as dissatisfied users often choose not to leave it~\cite{park2018positivity}. As a result, it can be ambiguous for workers seeking concrete guidance and insufficient for platforms monitoring quality.

These limitations are amplified in educational gig platforms. Learner feedback is often treated as a proxy for instructional quality, yet learner satisfaction does not always align with learning effectiveness. Meaningful learning often involves cognitive challenge, which can temporarily reduce satisfaction despite positive outcomes~\cite{carpenter2020students}. Thus, learner feedback alone may fail to give tutors actionable insights while limiting platforms' ability to maintain instructional standards.

To address these challenges, we explore AI-powered automated feedback as a complement. Since automated feedback is interpreted alongside high-stakes learner feedback in practice~\cite{wood2019good}, we investigate how tutors perceive and respond to it under this multisource context. We developed an automated feedback system as a research probe and deployed it on Ringle, a popular online English tutoring platform. The system analyzed tutors' lessons using the platform's internal evaluation standards and provided standardized, numerical feedback.

Our survey of 36 tutors found that, although tutors generally perceived automated feedback more negatively than learner feedback, they still found it valuable for self-monitoring and understanding platform expectations. Yet, conflicts with more lenient and high-stakes learner feedback caused confusion and frustration. Based on these findings, we propose design considerations for automated feedback systems in educational gig platforms at scale.
\section{Methods}

\subsection{Study Context} \label{studycontext}
Ringle\footnote{\url{https://www.ringleplus.com/}} is a popular online English tutoring platform offering 1-on-1 lessons with native English tutors. Since effectively leading tutoring sessions takes time to learn, tutors are trained through guidance documents and must pass Ringle's internal evaluation by conducting a mock lesson. After each lesson, Ringle encourages learners to leave feedback in three main types. First, learners can leave a \textit{5-star rating} on the lesson. Second, they can rate their \textit{willingness to meet the tutor again} on a 3-point Likert scale. Lastly, they can leave a \textit{free-form review} of the overall experience. The platform informs learners that 5-star ratings and free-form reviews are shared with the tutor (as feedback) and other learners (as reviews), while willingness-to-meet-again ratings are kept private and shared only with the platform so learners can express honest opinions.

\subsection{Research Probe}
We collaborated with three Ringle team members to develop automated feedback based on the platform's evaluation standards, previously used in mock lessons to familiarize tutors with internal expectations. These standards consisted of nine pedagogical categories (e.g., lesson structure, engagement), each assessed on a 5-point scale. To examine how tutors perceive standardized automated judgment, we developed an AI-powered feedback system as a research probe that prioritized transparency over complexity, reporting numerical scores for each of the nine subcategories generated from transcripts and tutors' notes~\cite{vitale2025noisy}. We used existing models with few-shot prompting~\cite{brown2020language}, and score thresholds were iteratively refined with the Ringle team until reaching satisfactory agreement with senior evaluators' manual ratings. Tutors received three reports during their first ten lessons, after the first, fifth, and tenth.

\subsection{Survey}

\subsubsection{Participants}
We launched the research probe to newly onboarded tutors\footnote{Tutors who joined Ringle within two weeks of the launch of the probe} on Ringle to examine perceptions of the automated feedback system without prior influence from learner feedback. After each tutor completed their first ten lessons, we distributed the survey. Participation was voluntary, and we excluded tutors who had not reviewed all ten learner feedback records or three automated feedback reports. A total of 36 tutors (ages 18--32; M = 23.28, SD = 3.41) participated. The survey took about 15 minutes, and each participant received a 10 USD Amazon gift card.

\subsubsection{Survey Questions and Analysis}
The survey had two sections. The first examined tutors' perceptions of automated feedback, with items adapted from prior work on evaluation systems~\cite{milanowski2001assessment,kelly2025high}. The same questions were asked for both automated and learner feedback, enabling direct comparison across seven dimensions: understanding, accuracy, fairness, favorableness, evaluator qualification, feedback uptake, and impact. Responses used a 7-point Likert scale (1=strongly disagree, 7=strongly agree). The second consisted of open-ended questions on tutors' challenges at Ringle, perceptions of subcategories, and suggestions for improvement. The full survey is available in the OSF Appendix\footnote{\url{https://osf.io/zsc3b/overview?view_only=dcf09d1a9fe64765b013df02af6bc382}}.

To compare tutors' perceptions between the two feedback sources, we conducted Wilcoxon signed-rank tests on the Likert responses. For open-ended responses, we conducted an inductive thematic analysis following the method of Braun and Clarke~\cite{braun2006using}.

\subsection{Feedback Data}
We gathered automated feedback from the first ten lessons of all 36 participants, including 327 of 360 lessons (33 were discarded due to recording issues). We also collected learner feedback from 10,000 randomly sampled lessons by 6,256 learners over a one-month period within the study duration. We then compared score distributions across the two sources to identify differences in how tutors were evaluated by the AI-powered automation system versus learners.

\section{Result}

\subsection{Perceptions of Automated Feedback} \label{survey}

\begin{figure}[hbt!]
    \centering
    \includegraphics[width=0.9\textwidth]{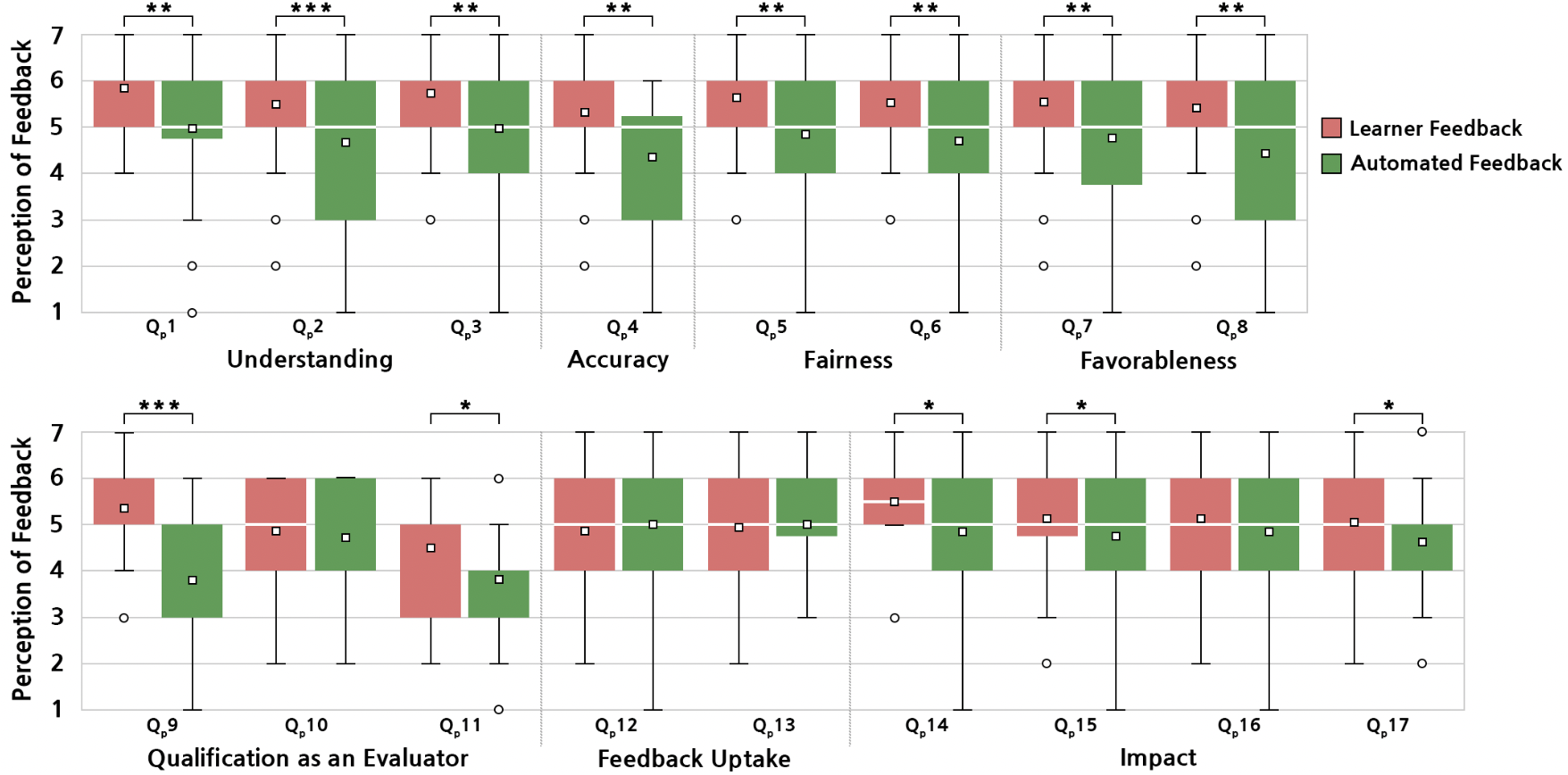}
    \caption{Tutors' perceptions of learner and automated feedback across understanding, accuracy, fairness, favorableness, qualification as an evaluator, feedback uptake, and impact (*\textit{p} < .05, **\textit{p} < .01, ***\textit{p} < .001)}
    \label{fig:perception}
\end{figure}

As shown in Figure~\ref{fig:perception}, tutors \textbf{perceived both learner and automated feedback positively overall}, with average ratings above 4 out of 7 across most dimensions, except for evaluator qualification for automated feedback. However, \textbf{tutors rated automated feedback significantly more negatively than learner feedback} across six dimensions: understanding, accuracy, fairness, favorableness, evaluator qualification, and impact. However, feedback uptake showed no significant difference.

From open-ended responses, tutors valued the automated feedback for providing \textbf{clear guidelines on the gig platform's expectations}, which helped them better understand performance standards (P3, P20, P19, P24, P28). They perceived this clarity as particularly useful for meeting platform requirements and adapting their teaching to align with structured evaluation standards.

\subsection{Differences Between Automated and Learner Feedback} \label{difference}

\begin{figure}[h]
    \centering
    \includegraphics[width=0.8\textwidth]{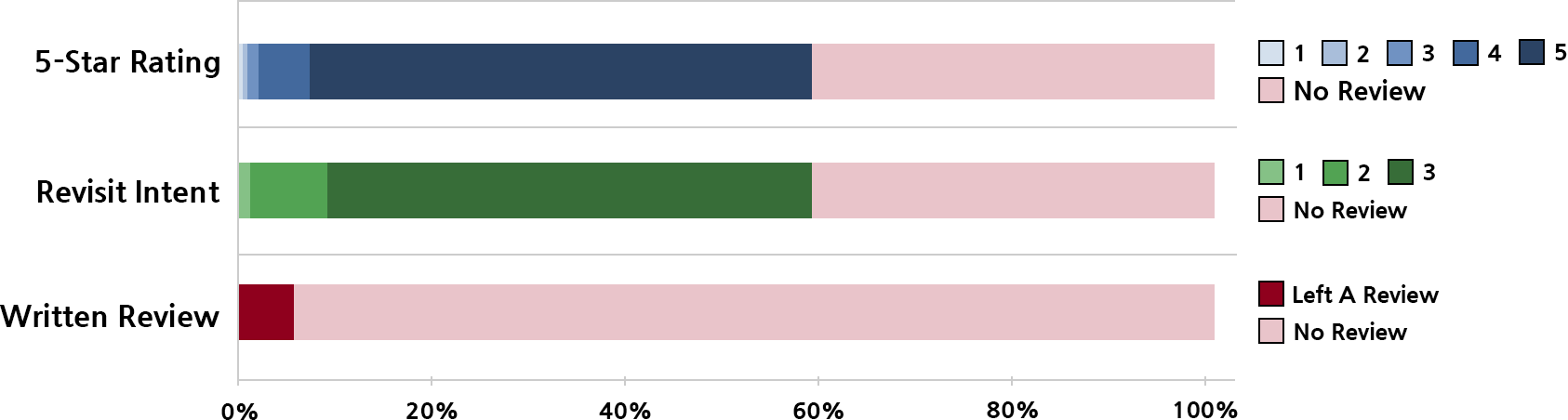}
    \caption{Proportion of learner feedback on 5-star ratings, revisit intent, and written reviews}
    \label{fig:learner}
\end{figure}

\begin{figure}[h]
  \centering
  \begin{subfigure}[t]{0.49\linewidth}
    \includegraphics[width=\linewidth]{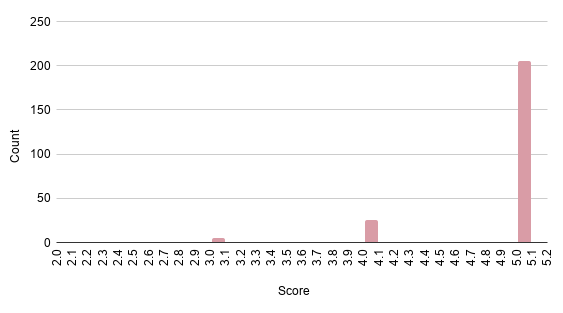}
  \end{subfigure}\hspace{0.5em}
  \begin{subfigure}[t]{0.49\linewidth}
    \includegraphics[width=\linewidth]{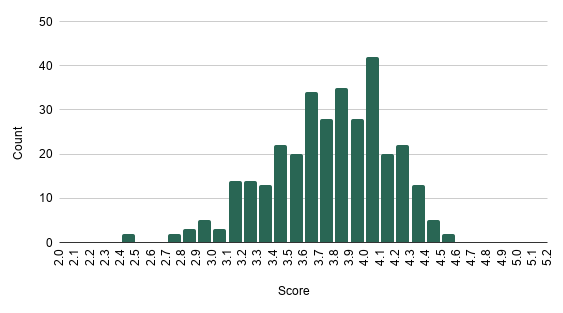}
  \end{subfigure}
  \caption{Histogram of 5-level scores from learner (left) and automated feedback (right)}
  \label{fig:histograms}
\end{figure}

As shown in Figure~\ref{fig:learner}, a substantial portion of lessons received \textbf{no learner feedback}, and when provided, it was \textbf{highly skewed toward positive ratings} (Figure~\ref{fig:histograms}, left). In contrast, automated feedback was consistently available and \textbf{showed a broader, more fine-grained distribution}, capturing greater variation in quality than discrete learner ratings (Figure~\ref{fig:histograms}, right).

\subsection{Responses to Feedback Differences} \label{responses}

\paragraph{Confusion Over Score Discrepancies.}
Many tutors \textbf{reported confusion over discrepancies between the two feedback sources}, particularly because both used comparable 5-point scales (P13, P18, P19, P27, P32, P34). Higher learner ratings (Figure~\ref{fig:histograms}) often led tutors to distrust automated feedback, consistent with prior findings on preferences for positive feedback~\cite{brett2001360}.

\paragraph{Contextual Gaps.}
Many tutors preferred learner feedback because they felt \textbf{the automated system could not account for varying lesson dynamics} (P4, P7, P25, P32, P35). Because the automated feedback lacked contextual awareness---such as learner requests, prior interactions, and time constraints---tutors perceived it as subjective. In contrast, learner feedback was favored as it came directly from the learners and was consequential in a high-stakes evaluative context.

\paragraph{Seeking Rationale.}
Tutors \textbf{wanted clearer explanations} to justify discrepancies between learner and automated feedback, since automated feedback only presented numerical scores without context (P3, P18, P33). This led them to feel that \textbf{automated feedback did not sufficiently reflect lesson contexts} (P2, P32, P34, P36) and \textbf{evaluations were inconsistent} (P2, P18, P24).

\paragraph{Self-Monitoring.}
Since learner feedback was highly skewed toward positive ratings, the \textbf{automated feedback became a valuable tool for self-monitoring and growth}. Tutors described using the automated feedback as a self-monitoring tool to identify areas for improvement, track their progress, and gain reassurance about their teaching effectiveness (P3, P35, P36).

\section{Design Considerations}

We discuss considerations for designing feedback systems in educational gig platforms.

\paragraph{DC1: Distinguish Feedback Types.}
When multiple feedback sources coexist in settings where learner feedback carries strong evaluative weight~\cite{wood2019good}, it is important to distinguish their purposes and scales. In our study, learner feedback tended to be positively skewed and was often missing altogether (\S\ref{difference}), limiting its usefulness as an evaluation tool. In contrast, automated feedback provided more consistent and critical assessments that supported tutors' self-monitoring (\S\ref{responses}). However, presenting feedback sources using similar scoring schemes led to confusion, creating a false impression of equivalent evaluations (\S\ref{responses}). Explicitly clarifying what each feedback measures and how to interpret the scores can help tutors make sense of multisource feedback.

\paragraph{DC2: Emphasize Complementary Nature.}
Learner and automated feedback should be framed as complementary rather than conflicting. While tutors often perceived automated feedback negatively---particularly when it failed to capture the lesson context (\S\ref{responses})---they found both types of feedback similarly helpful for planning future lessons (\S\ref{survey}). Discrepancies between feedback sources should therefore be presented as opportunities for reflection rather than signals to prioritize one source over another. Communicating which aspects are better assessed by learners (e.g., engagement) and which are more reliably captured by automated systems (e.g., lesson structure) can support tutors' interpretation and use of feedback for improvement.

\paragraph{DC3: Make Quality Standards Interpretable.}
Automated feedback can support gig platforms' quality control by organizing evaluations around structured criteria, rather than relying solely on learners' individual impressions. These criteria can make instructional standards more explicit and help tutors understand what the platform expects them to meet (\S\ref{survey}). To make these standards actionable, systems should pair scores with brief rationales, evidence, or contextual cues that explain how tutors can improve within platform expectations (\S\ref{responses}).

\section{Future Work}
Future work should examine how tutors' interpretations of automated feedback evolve over time, particularly as they gain experience navigating multiple feedback sources with unequal stakes. As automated feedback may function as a self-improvement tool rather than merely an evaluative signal, future studies could investigate how tutors adjust their teaching practices in response to such feedback and whether these adjustments lead to measurable improvements in lesson quality. It should also explore whether additional contextual or transparency cues meaningfully change how tutors weigh automated feedback relative to learner feedback in their teaching practices. 

Beyond tutors' perceptions, future work should examine the relationship between learner and automated feedback. This could include analyzing whether the two sources are directionally aligned when applied to the same lessons and how strongly they converge or diverge across feedback dimensions. Building on this analysis, future work could explore whether automated feedback can help calibrate positively skewed learner feedback. Understanding these relationships could inform the design of more interpretable feedback systems that help tutors make sense of multiple feedback sources while clarifying what each source is intended to capture.
\section*{Declaration on Generative AI}
In the preparation of this work, the authors used ChatGPT and Claude for grammar checking and improving the clarity of author-written text. The authors reviewed and revised all content and take full responsibility for the publication's content.

\bibliography{references}

\end{document}